\documentclass[aps,prd,preprint,groupedaddress,showpacs,showkeys,nofootinbib,draft]{revtex4-2}
\usepackage{bm,graphicx,amsmath}
\usepackage{amssymb}
\usepackage{amsmath}
\usepackage{bm}
\begin{document}
\title{The Intrinsic Angular - Momentum of Particles and the Resolution of the Spin-Statistics Theorem}
\author{Enrico Santamato}
\email{enrico.santamato@na.infn.it}
\affiliation{Dipartimento di Scienze Fisiche, Universit\`{a} di Napoli
    ``Federico II'', Italy}
\date{\today}
\author{Francesco De Martini}
\email{francesco.demartini2@gmail.com}
\affiliation{Accademia Nazionale del Lincei, Roma, Italy}

\newcommand{\christ}[2]{{#1 \brace #2}}
\newcommand{\bra}[1]{\ensuremath{\langle #1|}}
\newcommand{\ket}[1]{\ensuremath{|#1\rangle}}
\newcommand{\diag}{\ensuremath{\mathrm{diag}}}
\newcommand{\tint}{\textstyle\int}
\newcommand{\tiint}{\textstyle\iint}
\newcommand{\SO}{\ensuremath{\mathrm{SO}}}
\newcommand{\SU}{\ensuremath{\mathrm{SU}}}

\begin{abstract}
The traditional Standard Quantum Mechanics (SQM) theory is unable to solve the Spin-s problem, i.e., to justify the utterly important "Pauli Exclusion Principle". A complete and straightforward solution of the  Spin-Statistics problem is presented based on the "Weyl Integrable Quantum Mechanics" (WIQM) theory.  This theory provides a Weyl-gauge invariant formulation of the Standard Quantum Mechanics and reproduces successfully, with no restrictions, the full set of the quantum mechanical processes, including the formulation of Dirac's or Schr\"{o}dinger's equation, of Heisenberg's uncertainty relations, and of the nonlocal EPR correlations. etc. When the Weyl Integrable Quantum Mechanics is applied to a system made of many identical particles with spin, an additional constant property of all elementary particles enters naturally into play: the "intrinsic helicity", or the "intrinsic angular - momentum". This additional particle property, not considered by Standard Quantum Mechanics, determines the correct Spin-Statistics Connection (SSC) observed in Nature. All this leads to the consideration of a novel, most complete (in the EPR sense) quantum mechanical theory.
\end{abstract}
\maketitle
\section{Introduction}\label{sec:intro}
\textquotedblleft\textit{Everyone knows the Spin-Statistics theorem but no one understands it \dots The question is if physics contains this fact }[the Pauli principle]\textit{ as a prediction, and if so how this comes about; or whether physics is merely consistent with the Spin-Statistics Theorem and if some deeper explanation exists}\textquotedblright~\cite{duck99}. This puzzle represents the dramatic failure of the otherwise always successful Standard Quantum Mechanics (SQM) in a context of tremendous cosmological relevance since it involves the very existence of atoms, of ourselves, of the entire Universe. In the words of Richard Feynman: \textquotedblleft [It has been shown] \textit{that spin and statistics must necessarily go together, but we have not been able to find a way of reproducing his arguments on an elementary level. It appears to be one of the few places in Physics where there is a rule which can be stated very simply but for which no one has found a simple and easy explanation \dots This probably means that we do not have a complete understanding of the fundamental principles involved}\textquotedblright~\cite{feynman}. In the last decades, a vast literature has grown about the Spin-Statistics Connection (SSC) [see Refs .~\cite {duck98,duck99,romer02} for reviews and references]. In particular, attempts to model the quantum spin as a rotating frame attached to a point particle have appeared in the literature~\cite{balachandran93,jabs10,jabs10a}.\\

In the present Article, we shall present the explanation, in the nonrelativistic context, by introducing a novel, fundamental property of all elementary particles, \textquotedblleft intrinsic angular momentum\textquotedblright (IAM), as a necessary completion to the SQM. We show that the intrinsic angular momentum is a conserved quantity, implying a one-verse rotation constraint. The essential clue of the present demonstration resides in an insightful and EPR-complete description of the spin as due to the rotation of a frame described by three Euler angles, playing the role of nonlocal hidden variables~\cite{demartini14a}. Indeed, spin has always been considered by SQM and by Quantum Field Theory (QFT) to consist of an axial vector~\cite{debroglie} on the basis of a general understanding of the processes underlying all known experiments involving (mostly magnetic) spin interactions. We contend that the correct concept to be adopted for the spin lies down on its obvious and fundamental property as the angular momentum of a rotating frame attached to any specific particle and subjected to kinematic constraint, implied by first principles, which renders the particle similar to a ratchet gyroscope, where the rotation around its proper axis $\zeta$ can have only one sense (e.g. counter-clockwise). We show that this representation of the spin imposes an additional conserved, fundamental property to be attached to the particle, in addition to mass and charge: the \textquotedblleft intrinsic angular momentum\textquotedblright $s_\zeta$, i.e., the component of the angular momentum of the particle frame along its proper axis $\zeta$. This concept is absent in SQM. As shown by all standard texts of quantum mechanics, as in~\cite{davydov}, the concept of "intrinsic angular momentum" is extraneous to the lexicon of the SQM where one merely considers the three spin components $s_x, s_y, s_z$  along the axes $x,y,z$ of the laboratory frame. In the following, we will call \textquoteleft quantum spin\textquoteright\ the object described by the components $s_x, s_y, s_z$ (or their quantum operators) considered by SQM and proportional to the corresponding magnetic moments. The quantity $s_\zeta$ will be considered as a novel, independent physical quantity bearing several relevant, additional dynamical properties. For instance, although $s_\zeta$ has the same origin as $s_x, s_y, s_z$ (i.e., the rotation of the particle frame), it is a conserved, unmodifiable quantity whose value is not changed by any known force or potential acting on the particle. On the contrary, it is well known that $s_x, s_y, s_z$ can be modified by external fields so as to take any value of their spectrum from $-\hbar s$ to $+\hbar s$, with $s$ an integer or half-integer. Nevertheless, although quantum states with simultaneously fixed values of $s_x, s_y, s_z$ do not exist, owing to the noncommutativity of the corresponding operators, $s_\zeta$ retains its constant value together with each one of the spin components $s_x, s_y, s_z$ (the $s_\zeta$ operator commutes with all three spin operators). For simplicity and without loss of generality, we may orient the proper $\zeta$-axis of the particle frame to have $s_\zeta = +\hbar s$ and then refer to the constant spin value $s$ as an intrinsic "helicity" property of the particle, i.e., the \textquotedblleft intrinsic angular momentum\textquotedblright. The absence of $s_\zeta$ within the SQM structure justifies its failure to answer positively to the question reported at the beginning of the present Article. We find it impossible to fit the new helicity property coherently within the well-established structure of SQM. On the contrary, a straightforward solution to the SSC problem is attained in the context of the \textquotedblleft Weyl Integrable Quantum Mechanics\textquotedblright (WIQM), a modern Weyl gauge invariant approach to Quantum Mechanics~\cite{santamato12,santamato13,demartini14a,santamato84,santamato84a,santamato85,santamato88}. The WIQM was found to reproduce successfully all relevant processes of the SQM based on Dirac's or Schr\"{o}dinger's equations~\cite{santamato84,santamato13}, including Heisenberg's uncertainty relations~\cite{santamato88} and nonlocal EPR correlations of spin $\tfrac{1}{2}$ particles~\cite{santamato12,demartini14a,demartini14}. Within the WIQM perspective, the intrinsic angular momentum is found to fit in the theory coherently and naturally. For instance, the constraint of one rotation sense around $\zeta$ is not assumed \textit{ad hoc}, but emerges as a consequence of first principles. A fully relativistic version of the present SSC work is postponed to a forthcoming paper.

In Sec .~\ref{sec:onespin} we describe the single particle with spin, pointing out the main differences between the approaches of the WIQM and of the SQM;
in Sec.~\ref{sec:SSC} we present our proof of the Spin-Statistics Connection for $N$ identical particles with spin; and, finally in Sec.~\ref{sec:concl} we draw our conclusions.
\section{The single nonrelativistic particle with spin}\label{sec:onespin}
The WIQM assumes the existence of a frame attached to the particle, which is characterized not only by its location $\bm r=(x,y,z)$ in space but also by its orientation with respect to the fixed laboratory frame. We need three more parameters to describe orientation so that the configuration space of the particle is the six-dimensional space $E_3\times \SO(3)$. Any set of three parameters in the $SO(3)$ space can be chosen for the particle frame orientation, but the three Euler angles $\alpha,\beta,\gamma$ are proven to be the most convenient. We choose the six coordinates $q^i=\{x,y,z,a\alpha,a\beta,a\gamma\}$ with Euler angles so that $\alpha$ and $\beta$ are respectively the azimuthal and polar angle of the particle proper axis $\zeta$ in the fixed laboratory frame. The length $a$ can be considered as the "giration radius" of the particle, having mass $m$ and inertia momentum $I=ma^2$~\footnote{The giration $a$ can be assumed to be of the order of magnitude of the particle Compton length}. The kinetic energy $1/2 m v^2 + 1/2 I\omega^2$ induces in the  $E_3\times \SO(3)$ space a metric tensor $g_{ij}$ given by $\begin{pmatrix} I_3 & 0 \\ 0 & \Gamma_3\end{pmatrix}$, where $I_3$ is the 3D Euclidean metric, and $\Gamma_3$ is the $SO(3)$ metric $\begin{pmatrix}1 & 0 & \cos\beta \\
	0 & 1 & 0 \\
	\cos\beta & 0 & 1\end{pmatrix}$\\.

The main assumption of the WIQM is that this metric space is endowed with the nontrivial transport law for vectors given by Weyl's law $\delta\ell=-\ell\phi_idq^i$ with Weyl's vector $\phi_i=\frac{1}{n-2}\partial_i\rho$ where $n=6$ are the space dimensions. Our Weyl's connection is then \emph{integrable}, with Weyl's potential $\rho(q)$. The integrability condition ensures that Weyl's geometry is equivalent to Riemann's geometry up to a conformal factor, thus ruling out Einstein's "second clock effect" objection.\\

Weyl's transport law implies that our $E_3\times SO(3)$ space has a scalar curvature given by
\begin{equation}\label{eq:RW}
	R_W = \bar{R}-\xi^{-2}\frac{\nabla_k\nabla^k\sqrt{\rho}}{\sqrt{\rho}}
\end{equation}
where $n=6$ are the coordinate number, $\xi^2=\frac{n-2}{4(n-1)}=\frac{1}{5}$, and $\bar{R}=\frac{3}{2a^2}$, $\nabla_i$ are the Riemann scalar curvature and the covariant derivative calculated from the $E_3\times SO(3)$ metric $g_{ij}$, respectively. As noted elsewhere~\cite {santamato12}, the last term on the right of Eq.~(\ref{eq:RW}) turns out to be proportional to Bohm's quantum potential~\cite{bohm83a,bohm83b}. It is, therefore, natural to add to the kinetic energy the Weyl scalar curvature as the potential originating quantum phenomena. \\

The main equations of the WIQM are then given by~\cite{santamato84,santamato12,santamato13,demartini14a}
\begin{equation}\label{eq:HJE}
	-\partial_t S = \frac{1}{2m}g^{ij}\partial_i S\partial_j S + \frac{\xi^2\hbar^2}{2m}R_W
\end{equation}
\begin{equation}\label{eq:continuity}
	\partial_t\rho + \frac{1}{\sqrt{g}}\partial_i (\sqrt{g}g^{ij}\partial_j S) = 0
\end{equation}
where $R_W$ given by Eq.~(\ref{eq:RW}). \\
The \emph{ansatz} $\Psi=\sqrt{\rho}\exp(iS/\hbar)$ reduces the pair of Eqs.~(\ref{eq:HJE}) and (\ref{eq:continuity}) to the single Schr\"{o}dinger's equation 
\begin{equation}\label{eq:Schroedinger}
	i\hbar\partial_t\Psi = -\frac{\hbar^2}{2m}\left[\frac{1}{\sqrt{g}}\partial_i(\sqrt{g}g^{ij}\partial_j\Psi) - \xi^2\bar{R}\Psi\right]
\end{equation}
Because $\bar{R}$ is constant, the curvature term in Eq~(\ref{eq:Schroedinger}) can be eliminated by a suitable redefinition of the $\Psi$ phase.\\ 

The usual statistic interpretation of Quantum mechanics is restored by observing that any solution of the Hamilton-Jacobi Equation~(\ref{eq:HJE}) describes a bundle of geodesically equidistant trajectories in the $E_3\times SO(3)$ space, while the Continuity Equation~(\ref{eq:continuity}) provides a probability measure density $\rho$ over the bundle.
\\

The angle $\gamma$ is not present in the metric tensor, and if we assume $\rho$ independent of $\gamma$, this angle is not present in the Weyl curvature either. Therefore, $\gamma$ is an ignorable~\cite{taylor} (or cyclic~\cite{goldstein}) coordinate in the HJE~(\ref{eq:HJE}). We may then seek for solutions of Eqs.~(\ref{eq:HJE}) and (\ref{eq:continuity}) having the form
\begin{equation}\label{eq:S0}
    S(\alpha,\beta,\gamma;\bm r,t) = \hbar s \gamma + S_s(\alpha,\beta;\bm r,t)
\end{equation}
with constant $s$ and $\rho=\rho_s(\alpha,\beta;\bm r,t)$ only. We notice that $\rho_s$ and $S_s$ depend on $s$, because $s$ enters as a parameter in the reduced Hamilton-Jacobi equation obeyed by $S_s$ and in the corresponding continuity equation obeyed by $\rho_s$~\footnote{It can be shown that the addition of external e.m. fields introduces no dependence on $\gamma$ as well~\cite{santamato13,demartini14a}.}. The scalar wave function is then factorized as
\begin{equation}\label{eq:Psi}
   \Psi=e^{is\gamma}\Phi_s(\alpha,\beta;\bm r,t).
\end{equation}
with $\Phi_s=\sqrt{\rho_s}e^{iS_s/\hbar}$. We may then check \textit{a posteriori} that $\rho=|\Psi|^2=|\Phi|^2$ is independent of $\gamma$. It can be easily shown that $\hbar s_\zeta = \partial_\gamma S=\frac{1}{2I}\frac{d\gamma}{dt}$ is the component of the particle frame angular momentum along the particle proper $\zeta$-axis. We call $s_\zeta$ the intrinsic angular momentum of the particle, as said. Then Eq.~(\ref{eq:S0}) shows that the intrinsic angular momentum takes the constant value $s_\zeta=s$. No fields are known to be able to change the value of the intrinsic angular momentum, so once $s$ is given for a particular kind of particle, it is forever. We may thus consider the intrinsic angular momentum as a constant property of the particle, like charge and mass. Without loss of generality, we may assume that the rotation of the $\gamma$ angle around the particle $\zeta$-axis is counter-clockwise so to have
\begin{eqnarray}
	s &\ge& 0 \\
	\frac{d\gamma}{dt}&\ge& 0  \label{eq:gammadot}
\end{eqnarray}
The kinematic constraint (\ref{eq:gammadot}) shows that the spinning particle behaves as a sort of ratchet gyroscope with an internal mechanism preventing clockwise rotations around its proper axis (held fixed). Only counter-clockwise rotation is allowed.
As we shall see later, the kinematic constraint (\ref{eq:gammadot}) will have a crucial role in deriving the Spin-Statistic Connection.\\ 

In Eq.~(\ref{eq:S0}), $s$ can take any positive value, in principle. However, as we shall see below, the requirement of a normalizable measure, i.e. $\int\rho_s\sqrt{g}d^nq = 1$ and the properties of the unitary representations of the \SO(3) group imply that $s$ can take only integer or half-integer values so which coincides with the spin value of the particle in the SQM. \\

The SQM describes the state of a particle with spin $s$ by a $2s+1$-component spinor field $\psi^\sigma(\bm r,t)$ $(-s \le \sigma \le +s)$ defined in space and time only. The WIQM describes the same state with the scalar wave function $\Psi$ (or with the two fields $\rho$ and $S$) which depends on the angular variables. The connections between the two approaches is provided by the function $\Phi_s$ in Eq.~(\ref{eq:Psi}). In fact, the harmonic expansion in this case yields~\cite{santamato13,demartini14a}
\begin{eqnarray}\label{eq:Phi}
  \Phi_s(\alpha,\beta;\bm r,t)& = & \sum_{\sigma=-s}^s c_\sigma(\alpha,\beta)\psi^\sigma(\bm r,t)= \nonumber \\
                            & = & \sum_{\sigma=-s}^s e^{i\sigma\alpha} [d^{-1}(\beta)]^s_\sigma\psi^\sigma(\bm r,t)
\end{eqnarray}
where $s\equiv s_\zeta/\hbar$ is the particle intrinsic angular momentum, $c_\sigma(\alpha,\beta)=e^{i\sigma\alpha} [d^{-1}(\beta)]^s_\sigma$, and $d^s_\sigma(\beta)=D^s_\sigma(0,\beta,0)$ is the orthogonal part of the $(2s+1)$-dimensional unitary representation $D(\alpha,\beta,\gamma)$ of the rotation group. The unitary representations ensure that $\Phi_s$ is square summable (i.e. $\int\rho_s\sqrt{g}d^n q<\infty$) and restricts $s$ to be either an integer or half integer, as said above.\\

In the case of spin $\tfrac{1}{2}$, $\Phi_s$ reduces to
\begin{equation}\label{eq:Phidef}
   \Phi_s(\alpha,\beta;\bm r,t) = e^\frac{i\alpha}{2}\cos\frac{\beta}{2}\psi^\uparrow(\bm r,t)+e^{-\frac{i\alpha}{2}}\sin\frac{\beta}{2}\psi^\downarrow(\bm r,t)
\end{equation}
where $s=\tfrac{1}{2}$ and $\psi^\uparrow$ and $\psi^\downarrow$ are the components of the particle spinor with spin up and down with respect to the $z$-axis of the laboratory frame, respectively\footnote{The square modulus and phase of $\Phi_s$ are given by
\begin{eqnarray}
   |\Phi_s|^2 &=& \cos^2\frac{\beta}{2}|\psi^\uparrow(\bm r,t)|^2 + \sin^2\frac{\beta}{2}|\psi^\downarrow(\bm r,t)|^2 +\nonumber\\
                &&   \mbox{} +\cos\frac{\beta}{2}\sin\frac{\beta}{2}[\psi^{\uparrow*}(\bm r,t)\psi^\downarrow(\bm r,t)+\psi^\uparrow(\bm r,t)\psi^{\downarrow*}(\bm r,t)]\nonumber \\
   \arg(\Phi_s) &=& \arctan\left[\tan\frac{\alpha}{2}\left(\frac{\cos\frac{\beta}{2}\psi^\uparrow(\bm r,t)-\sin\frac{\beta}{2}\psi^\downarrow(\bm r,t)}{\cos\frac{\beta}{2}\psi^\uparrow(\bm r,t)+\sin\frac{\beta}{2}\psi^\downarrow(\bm r,t)}\right)\right]
\end{eqnarray}
}.
We note that the spinor $\psi^\sigma$ affects the function $\Phi_s$ only and not the phase factor $e^{is\gamma}$ in the scalar wave function (\ref{eq:Psi}). This phase factor is global and is ignored in the SQM, but in the WIQM is necessary to have $\Psi$ transforming as a scalar under rotation of the laboratory reference frame~\cite{santamato13,demartini14a}. Indeed, the scalar wave function $\Psi$ and the spinor $\psi^\sigma$ have different global properties. Under $2\pi$ rotation of the laboratory reference frame, the system is changed in itself, but, for half-integer spin, the spinor $\psi^\sigma$ changes in sign. On the contrary, the scalar wave function $\Psi$ of the WIQM remains invariant (it is a zero-spin field), as it can easily be checked. This is expected, however, because the action $S$ is invariant under $2\pi$ rotations. We see, therefore, that the WIQM handles fermions and bosons on equal footing in spite of their different physical behavior.\\

We observe, finally, that the connection between the WIQM and the SQM can be suitably stated and understood in terms of group theory. The factorization of the scalar wave function in Eq.~(\ref{eq:Psi}) parallels the factorization of the rotation group \SO(3) into a "boost" carrying the $z$-axis of laboratory frame into the particle proper $\zeta$-axis at polar angles $(\alpha,\beta)$, with no rotation around $\zeta$, and a successive rotation of an angle $\gamma$ around $\zeta$, viz.
\begin{equation}\label{eq:boost}
    R(\alpha,\beta,\gamma)=R_\zeta(\gamma)B(\alpha,\beta)=B(\alpha,\beta)R_z(\gamma),
\end{equation}
where $B(\alpha,\beta)=R(\alpha,\beta,0)=R_z(\alpha)R_y(\beta)$ denotes the "boost" operator\footnote{The terminology is taken from the similar factorization of the Lorentz group into rotations and boosts.}. The last equality in Eq.(\ref{eq:boost}) shows that the same orientation $R(\alpha,\beta,\gamma)$ can be obtained by first rotating the particle frame by an angle $\gamma$ around the laboratory $z$-axis and then applying the boost $B(\alpha,\beta)$. The angle $\gamma$ is thus defined as the angle of rotation around the $\zeta$-axis of the particle's proper frame as obtained from the laboratory frame by applying the boost $B(\alpha,\beta)$. From Eq.~(\ref{eq:boost}) we see that $\gamma$ parameterizes the rotation little group \SO(2) of rotations around the particle $\zeta$-axis. The angles $\alpha,\beta$ span, in turn, the quotient space \SO(3)/\SO(2) which is isomorphic to the sphere $S^2$. As we have seen, the function $\Phi_s$ in Eq.~(\ref{eq:Phidef}) is defined in this last space. Because $\Phi_s$ depends on the spinor $\psi^\sigma$ of the SQM, this picture is consistent with the usual description of the spinning particle as carrying a magnetic moment vector moving over the Bloch sphere. On the contrary, the WIQM describes the spin as a rotating frame, rather than a rotating vector. Frames differing from arbitrary rotation around the particle $\zeta$-axis are considered equivalent in the SQM, while they are considered different in the WIQM. As we shall see in the next Section, having extended the configuration space from the quotient space \SO(3)/\SO(2) of the SQM to the whole \SO(3) space of the WIQM is one of the key points to prove the SSC.
\section{The Spin-Statistics Connection}\label{sec:SSC}
Let us consider a system of $N$ particles. The configuration space is now $\bm X^N=\bm X\times\bm X\times\dots\times\bm X$, where $\bm X$ is the configuration space of each particle. If the particles are identical, a particle permutation leaves the system configuration unchanged, so that the configuration space is changed into $\bm X^N/\mathbb{S}^N$, where $\mathbb{S}^N$ denotes the group of permutations of $N$ objects. The Lagrangian of $N$ identical particles is invariant under $\mathbb{S}^N$ and so must be the action $S=\int Ldt$. Therefore we require, for any $p\in\mathbb{S}^N$,
\begin{equation}\label{eq:SNsymm}
   S(q_1,q_2,\dots,q_N,t) = S(p(q_1,q_2,\dots,q_N),t).
\end{equation}
As for one particle, also for $N$ particles with spin the angles $\gamma_a$ $(a=1,\dots,N)$ are ignorable in Eq.~(\ref{eq:HJE}). Because all particles have the same intrinsic angular momentum $s$, we may take
\begin{equation}\label{eq:SN}
   S(q_1,q_2,\dots,q_N,t) = \hbar s(\gamma_1+\dots+\gamma_N)+S_s(\tilde q_1,\dots,\tilde q_N,t).
\end{equation}
where $\tilde q$ denotes the coordinates with all angles $\gamma_a$ removed. Moreover, the field $\rho=|\Psi|^2$ obtained from Eq.~(\ref{eq:Psi}) with $S$ given by Eq.~(\ref{eq:SN}) is independent on the angles $\gamma_a$.\\

In the SQM the system of $N$ identical particles is described by a higher-order spinor $\psi^{\sigma_1,\dots,\sigma_N}(\bm r_1,\dots,\bm r_N,t)$. The connection with the WIQM is made through the scalar wave function $\Psi$ that, in this case, becomes
\begin{equation}\label{eq:PSiN}
   \Psi =\Psi(\alpha_1,\beta_1,\gamma_1, \bm r_1;\dots,\alpha_N,\beta_N,\gamma_N,\bm r_N,t) =e^{is\sum_{a=1}^N \gamma_a}\Phi_s(\tilde q_1,\dots,\tilde q_N,t)
\end{equation}
where $\Phi_s$ is related to the spinor $\psi^{\sigma_1,\dots,\sigma_N}$ by
\begin{eqnarray}\label{eq:PhiN}
   \lefteqn{\Phi_s(\alpha_1,\beta_1,\bm r_1;\dots;\alpha_N,\beta_N,\bm r_N,t) =} \nonumber \\
     &=& \sum_{\sigma_1=-s}^s\dots\sum_{\sigma_N=-s}^s c_{\sigma_1}(\alpha_1,\beta_1)\dots c_{\sigma_N}(\alpha_N,\beta_N)\times \nonumber\\
     & &\mbox{} \times \psi^{\sigma_1\dots\sigma_N}(\bm r_1,\dots{\bm r}_N,t)
\end{eqnarray}
which generalizes Eq.~(\ref{eq:Phi}) to $N$ identical particles with intrinsic angular momentum $s$.\\

Let us consider now the exchange of the coordinates of two particles $a$ and $b$, say, while the coordinates of all other particles remain fixed. To replace the angles $(\alpha_a,\beta_a,\gamma_a)$ with $(\alpha_b,\beta_b,\gamma_b)$ we apply to the frame of particle $a$ the rotation operator $R_{a\rightarrow b}=R(\alpha_b,\beta_b,\gamma_b)R^{-1}(\alpha_a,\beta_a,\gamma_a)$. The exchange of the angles is completed by applying to the frame of particle $b$ the rotation $R_{b\rightarrow a}$ obtained from $R_{a\rightarrow b}$  by exchanging the suffix $a$ with $b$. In these operations, the space coordinates $\bm r_a,\bm r_b$ of the two particles remain fixed. Then, from Eq.~(\ref{eq:boost}) we find
\begin{eqnarray}\label{eq:Rab}
   R_{a\rightarrow b}&=& B(\alpha_b,\beta_b)R_z(\gamma_b-\gamma_a)B^{-1}(\alpha_a,\beta_a) \nonumber \\
   R_{b\rightarrow a} &=& B(\alpha_a,\beta_a)R_z(\gamma_a-\gamma_b)B^{-1}(\alpha_b,\beta_b)  
\end{eqnarray}
so that the angle of exchange rotation is given by
\begin{equation}\label{eq:Rabba}
   R_{a \leftrightarrow b}=R_{a\rightarrow b}R_{b\rightarrow a}=B(\alpha_b,\beta_b)R_z(\gamma_b-\gamma_a)R_z(\gamma_a-\gamma_b)B^{-1}(\alpha_b,\beta_b)
\end{equation}
The factorization (\ref{eq:boost}), implies that the angular coordinates of particles $a$ and $b$ can be exchanged in the following steps: first we use boosts $B^{-1}(\alpha_a,\beta_a)$ and $B^{-1}(\alpha_b,\beta_b)$ to align the axes $\zeta_a$ and $\zeta_b$ of the frames of the two particles along the common direction of the laboratory $z$-axis. Then, we  exchange the angles $\gamma_a$ and $\gamma_b$ by rotating around $z$ and, finally, we reset the $\zeta$-axes in their exchanged orientations applying the boosts $B(\alpha_a,\beta_a)$ and $B(\alpha_b,\beta_b)$.\\

The pivotal point is that in exchanging the $\gamma$-angles clockwise rotations around the fixed $z$-axis are forbidden by the constraint $d\gamma/dt\ge 0$ [see Eq.~(\ref{eq:gammadot})], so that the fixed axis rotation product $R_z(\gamma_a-\gamma_b)R_z(\gamma_b-\gamma_a) = R_z(2\pi) =1)$, as due because the frame orientation is restored to its previous configuration. Nevertheless, the change of the particles $\gamma$-angles is $\Delta\gamma_a=\gamma_b-\gamma_a= 2\pi-\Delta\gamma_b$. It follows that the change of the sum $\gamma_a+\gamma_b$ is given by $\Delta\gamma_a+\Delta\gamma_b = 2\pi$. \\

When this result si applied to Eq.~(\ref{eq:SN}), the angular coordinates are exchanged, but the action $S$ is incremented of $2\pi\hbar s$ and the scalar wave function $\Psi$ is multiplied by the phase factor $e^{2i\pi s}=(-1)^{2s}$~\footnote{Had we chosen clockwise rotation around the particle proper $\zeta$ axis, we would have obtained in all formulae above $-2\pi$ in place od $2\pi$, but the wavefunction phase factor would be equally unchanged.}. Because any permutation $p\in\mathbb{S}^N$ is obtained by a finite number $k_p$ of simple transpositions, we may apply the permutation $p$ to the angular coordinates of the action $S$ by applying $k_p$ exchanges of the angular coordinates of two particles at a time as described above. Each one of these exchanges add a term $2\pi\hbar s$ to $S$ so that, after the permutation $p$ of the angular coordinates, $S$ is incremented by $\Delta S = 2\pi k_p\hbar s$ and $\Psi$ is multiplied by $(-1)^{2k_p s}$. This is in contrast with the general symmetry requirement (\ref{eq:SNsymm}) for the action.\\

Since now, however, only the angular coordinates have been permuted. In order to restore the symmetry (\ref{eq:SNsymm}), among all solutions of Eqs.~(\ref{eq:HJE}) and (\ref{eq:continuity}) (or the corresponding wave equation for $\Psi$), we must peek only the ones which, under the permutation $p$, applied to the space coordinates only, compensates the additional term $\Delta S= 2\pi k_p\hbar s$ in the action $S$ (or the corresponding phase factor $(-1)^{2k_p s}$ in the scalar wave function $\Psi$).\\

We now observe that, because the $\sigma$ are dummy indices in Eq.~(\ref{eq:PhiN}), any permutation of the coordinates in the function $\Phi_s$ on the right of this equation is equivalent to the same permutation of the particle \textit{labels} (i.e. to the simultaneous permutation of coordinates \emph{and} spin) in the term in the sum on the right. Therefore, the only possible choice for the spinor $\psi^{\sigma_1,\dots,\sigma_N}$ in Eq.~(\ref{eq:PhiN}) is to take
\begin{eqnarray}\label{eq:psisymm}
  \lefteqn{\psi^{\sigma_1,\dots\sigma_N}(\bm r_1,\dots,\bm r_2,t) =} \nonumber \\
      &=& \frac{1}{N!}\sum_{\alpha=1}^{N!}(-1)^{2s}p_\alpha[\psi_1^{\sigma_1}(\bm r_1,t)\dots \psi_N^{\sigma_N}(\bm r_N,t)]
\end{eqnarray}
where the sum is extended to all $N!$ permutations $p_\alpha\in\mathbb{S}^N$ of the one-particle spinor states, in accordance with Pauli's Spin-Statistics Connection. Parastatistics are automatically excluded.
\section{Conclusions}\label{sec:concl}
In conclusion, we presented a proof of the SSC in the nonrelativistic limit without making recourse to the quantum field approach. In the WIQM framework, this is made possible by the role played by the intrinsic angular momentum $s_\zeta$  and by the peculiar kinematical constraint $d\gamma/dt\ge 0$, arising from the conservation of $s_\zeta$. We may indeed regard the intrinsic angular momentum as a constant property of any elementary particle, which is not considered in the SQM. In the WIQM, the SSC theorem is a consequence of the allowed values of $s_\zeta$, of the kinematic constraint (\ref{eq:gammadot}), and of the general requirement (\ref{eq:SNsymm}) to have an action $S$ invariant under permutation of the angular and space coordinates. Most importantly, it follows that the WIQM handles bosons and fermions on the same footing by a unique scalar wave function $\Psi$. It is the space-time part of this wave function, contained in the spinor fields inside the reduced wave function $\Phi_s$, which behaves differently under exchange of identical particles, according to Pauli's principle. Because the SSC is related to the particle rotational properties, the SSC demonstration runs equally in the case of relativistic particles, provided the WIQM approach is used, as made, for instance, in the case of Dirac's particle \cite{santamato17}.  In summary, the achievement of the \textquotedblleft intrinsic angular momentum\textquotedblright WIQM theory in the present, utterly critical domain of the SSC can be extended, with no restrictions, to the full set of physical phenomena and legitimize the consideration of the \textquotedblleft intrinsic angular momentum\textquotedblright WIQM as the most complete quantum mechanical theory of general applicability. \\


\bf{References}\\
%
\end{document}